\documentstyle[preprint,prd,aps,psfig]{revtex}\tighten
\newcommand{\beq}{\begin{equation}}
\newcommand{\eeq}{\end{equation}}
\newcommand{\beqa}{\begin{eqnarray}}
\newcommand{\eeqa}{\end{eqnarray}}

\newcommand{\siml}{\lesssim}

\newcommand{\k}{\kappa}
\newcommand{\vp}{\varphi}

\def\CQG#1#2#3{Class. Quantum Grav. {\bf #1}, #2 (19#3)}

\def\JMP#1#2#3{J. Math. Phys. {\bf #1}, #2 (19#3)}

\def\PLB#1#2#3{Phys. Lett. B {\bf #1}, #2 (19#3)}
\def\PLBB#1#2#3{Phys. Lett. B {\bf #1}, #2 (20#3)}
\def\PLBold#1#2#3{Phys. Lett. {\bf#1B}, #2 (19#3)}
\def\PRD#1#2#3{Phys. Rev. D {\bf #1}, #2 (19#3)}
\def\PRDD#1#2#3{Phys. Rev. D {\bf #1}, #2 (20#3)}

\def\PRLL#1#2#3{Phys. Rev. Lett. {\bf#1}, #2 (20#3)}

\begin{document}
\widetext
\draft
\preprint{\tighten\vbox{\hbox{KUNS-1859}\hbox{astro-ph/0307338}}}

\title{
$1/R$ Gravity and Scalar-Tensor Gravity
}

\author{Takeshi Chiba}
\address{
Department of Physics, Kyoto University, Kyoto 606-8502, Japan
}

\date{\today}

\maketitle

\begin{abstract}
We point out that extended gravity theories, the Lagrangian of which is 
an arbitrary function of scalar curvature $R$, are equivalent to 
a class of the scalar tensor theories of gravity. 
The corresponding gravity theory is $\omega=0$ Brans-Dicke gravity with a 
potential for the Brans-Dicke scalar field, 
which is not compatible with solar system experiments if the field 
is very light: the case when such modifications become important 
recently. 
\end{abstract}

\vspace{1.5cm}

The problem of dark energy is the problem of 
$\Omega$: $\Omega=8\pi G\rho_M/3H^2 <1$. Since $\Omega$ can be 
regarded as the ratio of the right-hand-side of the Einstein equation 
(matter) to the left-hand-side of the Einstein equation (curvature=gravity), 
in order to make $\Omega=1$ one requires either (i) introduction of new 
form of matter(energy): dark energy or (ii) modification of gravity in the 
large, so that the total energy density is equal to the critical density, 
which is required by theory (inflation) or by observation (WMAP). 

Recent attempts to modify gravity by introducing $R^{-1}$ 
term \cite{cdtt,cct} fall in the latter possibility:
\footnote{If such models are phenomenologically viable, 
$R^{-1}$ gravity might be called ``c-essence'' (c for curvature).} 
\beq
S={1\over 2\k^2}\int d^4x\sqrt{-g}\left(R-{\mu^4\over R}\right)+
S_{matter}(g_{\mu\nu}),
\label{action1}
\eeq
where  $\k^2=8\pi G$. 
The Newtonian limit of such modified gravity theories is studied in 
\cite{dick}, and it is found that Newton gravity is reproduced (as it 
should be). In this note, we point out that modified gravity theories with 
the Lagrangian of an arbitrary function of $R$ are equivalent to 
a special class of scalar tensor theories of gravity. 
We also calculate the PPN (parameterized post-Newtonian) parameter 
of such gravity theories. To this end, we utilize the dynamically equivalent 
action by introducing a new field $\phi$ \cite{no} :
\beq
S={1\over 2\k^2}\int d^4x\sqrt{-g}\left(\left(1+{\mu^4\over \phi^2}\right)R-
{2\mu^4\over \phi}\right)+S_{matter}(g_{\mu\nu}).
\eeq
One can easily verify that the field equation for $\phi$ gives $\phi =R$, 
which reproduces the original action Eq.(\ref{action1}).
\footnote{The field $\phi$ is not a auxiliary field since 
the field equations contain the second derivative of $\phi$ through the 
equation of motion of the metric.} 

The equivalence is easily generalized to arbitrary function of $R$:
\beq
S={1\over 2\k^2}\int d^4x\sqrt{-g}F(R)+S_{matter}(g_{\mu\nu}).
\eeq
The equivalent action is \cite{tt,wands}:
\beq
S={1\over 2\k^2}\int d^4x\sqrt{-g}\left(F(\phi)+F'(\phi)(R-\phi)\right)+
S_{matter}(g_{\mu\nu}),
\label{action2}
\eeq
where $F'(\phi)=dF/d\phi$. 
One can easily verify that the field equation for $\phi$ gives $\phi =R$ 
if $F''(\phi)\neq 0$, which reproduces the original action.
After the conformal transformation such that 
$F'(\phi)g_{\mu\nu}=g_{\mu\nu}^E$ along $\phi=R$, 
the action is reduced to that of the 
scalar field minimally coupled to the Einstein gravity\cite{whitt,bc,maeda,ms}:
\beqa
S&=&{1\over 2\k^2}\int d^4x\sqrt{-g_E}\left(R_E-
{3\over 2F'(\phi)^2}g_E^{\mu\nu}
\nabla_{E\mu}F'(\phi)\nabla_{E\nu}F'(\phi)-{1\over F'(\phi)^2}
\left(\phi F'(\phi)-F(\phi)\right)
\right)\nonumber\\
&&+S_{matter}(g^E_{\mu\nu}/F'(\phi)).
\label{action3}
\eeqa
Introducing a canonical scalar field $\vp$ such that 
$F'(\phi)=\exp(\sqrt{2/3}\k\vp)$, Eq.(\ref{action3}) can be written as 
\beqa
S&=&\int d^4x\sqrt{-g_E}\left({1\over 2\k^2}R_E-{1\over 2}(\nabla\vp)^2
-V(\vp)\right)+S_{matter}(g^E_{\mu\nu}/F'(\phi(\vp))),\\
&&V(\vp)=\left(\phi(\vp) F'(\phi(\vp))-F(\phi(\vp))\right)/
2\k^2F'(\phi(\vp))^2.\nonumber
\eeqa

So the question arises: what is the gravity described by the original 
frame metric $g_{\mu\nu}$? Since the gravity described by $g_{\mu\nu}^E$ 
is the Einstein-scalar system and $g_{\mu\nu}(=g^E_{\mu\nu}/F'(\phi))$ is 
admixture of spin 0 degree of freedom and spin 2 degree of freedom, 
the gravity by $g_{\mu\nu}$ should be a class of scalar-tensor theories 
of gravity which are subject to observational constraints coming from 
the solar system experiments of gravity \cite{will}. 
Usually higher derivative modifications of gravity are thought to be important 
in the early universe, and hence the bounds on $\omega$ by the present 
time experiments are not important. However, if such modifications become 
important recently, there is the danger that such theories may be in conflict 
with experiments. 
In fact, the absence of the kinetic term in Eq.(\ref{action2}) implies that 
the Brans-Dicke parameter is vanishing, $\omega=0$ 
(or the PPN parameter $\gamma=(\omega +1)/(\omega +2)$ is $\gamma=1/2$). 
The current bound on $\omega$ is $\omega>3500$ 
(or $|\gamma-1|<2.8\times 10^{-4}$) \cite{will}. 
This bound applies to the Brans-Dicke type 
theory with the very light Brans-Dicke scalar field 
with mass $\siml$ (1 A.U.)$^{-1}\sim 10^{-27}$GeV (e.g. extended quintessence)\cite{chiba}.

We estimate the effective mass for two examples. The first example is 
the Starobinsky model \cite{alex}: $F(R)=R+R^2/M^2$ with $M\sim 10^{12}$GeV. 
In terms of the scalar field $\vp$, the effective potential can be 
rewritten as 
\beq
V(\vp)={M^2e^{-2\sqrt{2/3}\k\vp}\over 8\k^2}
\left(e^{\sqrt{2/3}\k\vp}-1\right)^2,
\eeq
where we have neglected the matter term for simplicity. Evaluating 
the second derivative of $V(\vp)$ \footnote{Note that  
$3d^2V/d\vp^2=1/F''+\phi/F'-4F/F'^2$.} 
around the Minkowski vacuum ($\vp=0$) 
gives the effective mass squared of the scalar field of order $M^2$, 
which is much larger than $H_0^2$. Hence the constraints by the 
solar system experiments do not apply here.

The second example is CDTT model \cite{cdtt,cct}: $F(R)=R-\mu^4/R$ 
with $\mu\sim H_0\simeq 10^{-42}$GeV. Again, in terms of $\vp$, 
the effective potential is given by
\beq
V(\vp)={\mu^2e^{-2\sqrt{2/3}\k\vp}\over \k^2}\sqrt{e^{\sqrt{2/3}\k\vp}-1}.
\eeq
Evaluating $V''$ around $\phi=R\sim H_0^2$ $(\k\vp\sim 1)$ gives 
the effective mass squared of order $\mu^2$ (and 
tachyonic for $8/9<e^{\sqrt{2/3}\k\vp}<2$), which is very light. 
Together with $\omega=0$, the solar system experiments exclude such a 
theory.\footnote{If we can create a dip in $V(\vp)$ at $\vp\sim \mu$ 
so that $V''\gg \mu^2$ there (like Albrecht-Skordis model \cite{as}), then 
we may evade the constraints. However, we are currently unable to 
construct such $F(R)$. } 

To conclude, we have shown that extended gravity theories, the Lagrangian 
of which is an arbitrary function of scalar curvature $R$, are equivalent to 
a class of the Brans-Dicke type theories of gravity with a potential. 
The corresponding Brans-Dicke parameter is $\omega=0$. If such 
modifications become important recently, the scalar field is generically 
very light and mediates a gravity force of long range. 
Hence such theories are not compatible with solar system experiments.  
Thus c-essence may cease to exist. It remains to be seen whether other 
modification of gravity (higher dimensional origin \cite{dgp}, 
massive graviton \cite{dkp}, etc) 
could be phenomenologically viable alternative to dark energy. 

{\it Note added:} Ref. \cite{dk} addresses the stability issue of 
Eq.(\ref{action1}). 

\acknowledgments

We would like to thank Takahiro Tanaka for useful discussion. 
This work was supported in part by a Grant-in-Aid for Scientific 
Research (No.15740152) from the Japan Society for the Promotion of
Science and by a Grant-in-Aid for Scientific Research on Priority Areas 
(No.14047212) from the Ministry of Education, Science, Sports and 
Culture, Japan. 


\end{document}